\documentclass[]{emulateapj}

\begin{document}
\slugcomment{Submitted to ApJ Letters 2009 June 12}
\shorttitle{Discovery of H$_2$* near GRB 080607}
\shortauthors{Sheffer et al.}

\title{The Discovery of Vibrationally-Excited H$_2$ in the Molecular Cloud
near GRB 080607}
\author{Y. Sheffer\altaffilmark{1},
	J. X. Prochaska\altaffilmark{2,3},
 	B. T. Draine\altaffilmark{4},
	D. A. Perley\altaffilmark{5},
	and J. S. Bloom\altaffilmark{5}}
\altaffiltext{1}{Department of Physics and Astronomy,
University of Toledo, Toledo, OH 43606; ysheffe@utnet.utoledo.edu}
\altaffiltext{2}{University of California Observatories/Lick Observatory,
University of California, Santa Cruz, CA 95064} 
\altaffiltext{3}{Department of Astronomy and Astrophysics,
University of California, Santa Cruz, Santa Cruz, CA 95064} 
\altaffiltext{4}{Princeton University Observatory,
Peyton Hall, Ivy Lane, Princeton, NJ 08544}
\altaffiltext{5}{Department of Astronomy, University of California,
Berkeley, CA 94720-3411}

\begin{abstract}

GRB 080607 has provided the first strong observational signatures of molecular
absorption bands toward any galaxy hosting a gamma-ray burst.
Despite the identification of dozens of features as belonging to
various atomic and molecular (H$_2$ and CO) carriers, many more
absorption features remained unidentified.
Here we report on a search among these features for absorption from
vibrationally-excited H$_2$, a species that was predicted
to be produced by the UV flash of a GRB impinging on a molecular cloud.
Following a detailed comparison between our spectroscopy
and static, as well as dynamic, models of H$_2$* absorption,
we conclude that a column density of 10$^{17.5\pm0.2}$
cm$^{-2}$ of H$_2$* was produced along the line of sight toward GRB 080607.
Depending on the assumed amount of dust extinction between the molecular
cloud and the GRB, the model distance between the two is found to be
in the range 230--940 pc.
Such a range is consistent with a conservative lower limit of 100 pc estimated
from the presence of \ion{Mg}{1} in the same data.
These distances show that substantial molecular material is found
within hundreds of pc from GRB 080607, part of the distribution of clouds
within the GRB host galaxy.

\end{abstract} 

\section{Introduction}

Long-duration GRBs are currently understood to be the electromagnetic 
manifestation of highly beamed energy from the sites of the demise of
young massive stars \citep[see, e.g.,][]{Woos06}.
This implies that GRB progenitors must have formed and lived near molecular
clouds in their respective host galaxies.
Direct evidence for the molecular nature of their birthplace has been lacking,
presumably owing to complete destruction of molecules within $\sim$100 pc of
the GRB \citep{tpc+07,Whal08}.
Recently, \citet[][hereafter P09]{Proc09} presented the first unambiguous
spectroscopic detection of absorption bands from H$_2$ and CO molecules
along a translucent GRB sight line (GRB~080607, z = 3.0363).
Their initial analysis provided the first discovery of molecular gas with
properties very similar to those of Galactic molecular clouds, but
located in the star-forming ISM of a GRB host galaxy.

The detection by P09 of neutral species (e.g., \ion{C}{1}, \ion{Mg}{1}) toward
GRB 080607 indicates that the bulk of the atomic gas was $\geq$100 pc from
the GRB, presumably beyond the photoionization sphere formed by the
progenitor \citep{Whal08}.
However, whereas \citet{Whal08} introduced an additional galactic-wide
FUV field as means to suppress the formation of unobserved H$_2$, the discovery
of more than 10$^{21}$ cm$^{-2}$ of H$_2$ toward GRB 080607, well shielded
by dust extinction (rest-frame $A_V = 3.2$ mag), raises the possibility that
the molecular material could have been in close proximity to the GRB.
The detection by P09 of absorption from rotationally-excited
CO, and the determination that atomic and molecular lines are kinematically
distinguished by
30 $\pm$ 15 km s$^{-1}$, specifically allow for this scenario.

\citet{Drai00}, followed by \citet[][hereafter DH02]{Drai02}, predicted that
observable quantities of vibrationally-excited H$_2$ (from $v'' > 0$,
hereafter H$_2$*) can be produced by UV photons in molecular gas
within 100 pc from a GRB source.
With an assumed model peak luminosity of $L_0$ = 1/40 of that observed
for GRB 990123,
DH02 produced a column density ($N$) of $\approx$10$^{19}$ cm$^{-2}$
of H$_2$* in a gas with an initial $N$ = 10$^{21}$ cm$^{-2}$
of cold H$_2$ at close proximity ($\leq$1 pc) to the GRB.
P09 reported cold H$_2$ with $N \approx 10^{21.2}$ cm$^{-2}$ toward GRB 080607,
suggesting the possibility that H$_2$* could also have been present.

Here we report on our successful search for absorption from H$_2$*
toward GRB 080607.
Previously, one Galactic sight line was shown to have $\sim$500
absorption lines associated with H$_2$* \citep[][toward HD 37903]{Meye01},
while the sample of low-redshift galaxies, meanwhile, that are found to emit
ro-vibrational H$_2$* lines near 2.1 $\micron$ has been increasing steadily,
albeit mostly attributed to shocked gas \citep{Thom78,Jaff97,Dona00,Dale09}.
Our sight line toward GRB 080607 at the redshift of z = 3.0363
is thus the most distant detection of H$_2$*, as well as the first
extra-galactic detection of H$_2$* via absorption:
this was made possible thanks to an extremely bright GRB, an extremely
capable optical system (the LRIS spectrometer on the Keck~I telescope),
and a rapid response to the GRB alert from $Swift$.
We also show, by a comparison with theoretical predictions, that this
detection serves to confirm the general picture of the interaction
between a GRB and its immediate galactic environment.

\begin{figure*}
\begin{center}
\includegraphics[height=6.8in,angle=90]{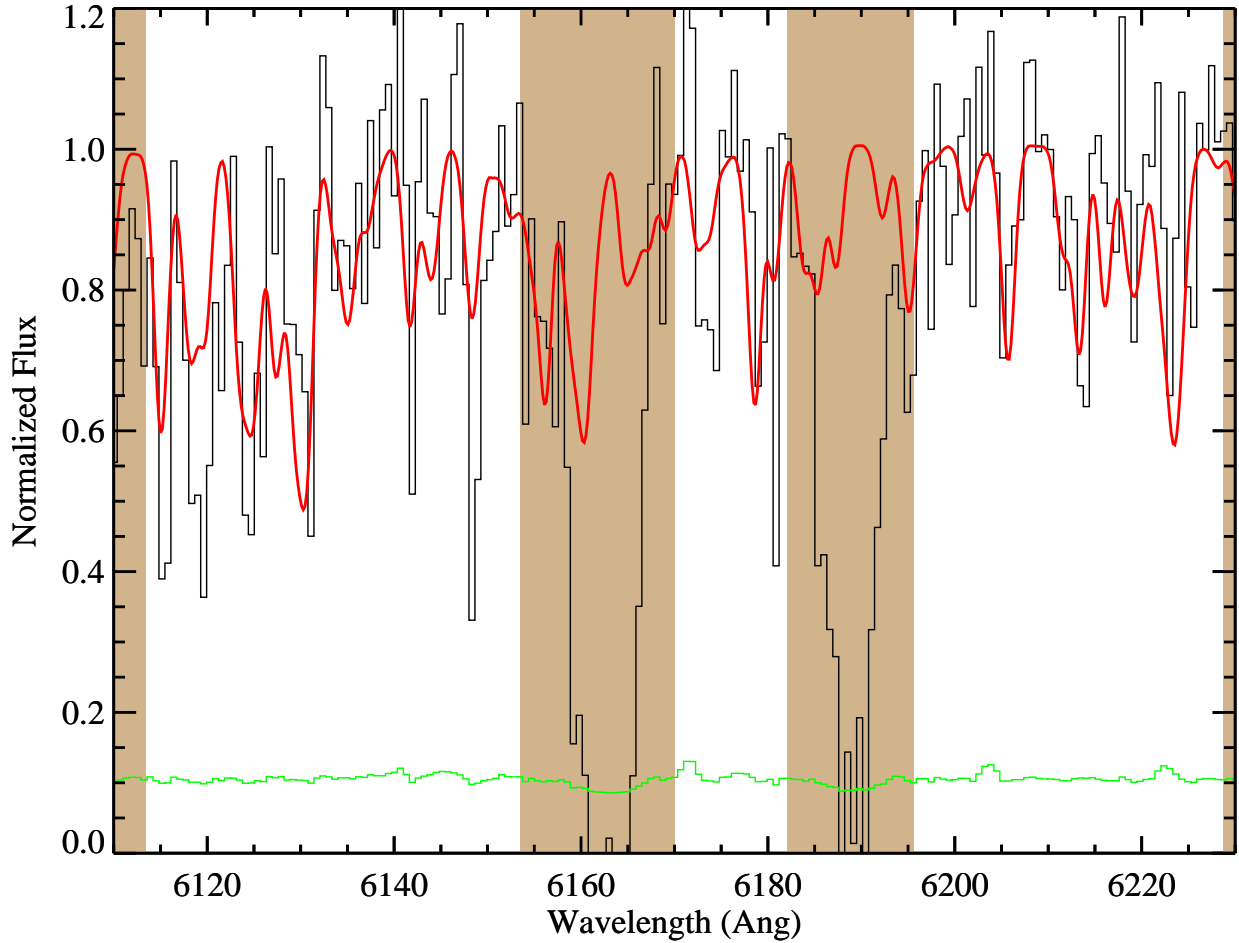}
\end{center}
\caption{Zoomed comparison between the Keck/LRIS R1200
data (black histogram) and the H$_2$* dynamic model (smooth red line).
Brown masking designates absorption from other identified species
(e.g., \ion{Si}{2} at 6163 and 6189 \AA),
where the detection of H$_2$* would be compromised by blending.
The agreement is very good between stronger H$_2$* absorption features and
equally strong, hitherto unidentified ``lines'' in the spectrum.
The noise level is denoted by the green line at the bottom.}
\label{fig:r1200_zoom}
\end{figure*}

\section{Observations and Modeling}

The acquisition of the spectroscopic data analyzed here was described by P09.
Here we recap some essential information relating to the sequence of events
in the rest frame of the GRB, owing to the predicted temporal evolution
of the H$_2$* abundance (DH02).

GRB 080607 triggered the Swift detector at 06:07:27 UT on 2008 June 7
\citep{Mang08}, here defined as $t$ = 0 s.
A first series of three Keck/LRIS exposures was taken by us (DAP and JSB)
with the B600 and R400 gratings, simultaneously employing the blue
and red cameras.
These exposures were 75, 150, and 300 co-moving seconds long, centered on
rest-frame times $t$ = 340, 490, and 730 s.
A second series of two exposures was taken with the B600 and R1200
gratings, both 370 s long, and centered on the post-trigger times $t$ = 1230
and 1640 s in the rest frame of the GRB.
In terms of spectral resolution, the sub-arcsecond seeing resulted
in $\lambda$/$\Delta$$\lambda$ = $R$ $\sim$ 4000, 2000, and 1000, for
gratings R1200, B600, and R400, respectively.

Our initial search for possible H$_2$* features involved spectral syntheses
of relevant transitions with the Y.S. code Ismod.f, and
comparisons with the highest-$R$ data from the R1200 grating.
Each transition was modeled with a single Voigt profile,
employing as fixed parameters the resolution $R$, transition rest
wavelength ($\lambda_{\rm rest}$) and oscillator strength ($f$-value),
the Doppler parameter $b$, and a radial velocity of +30~km s$^{-1}$
relative to the atomic gas (P09).
Both the $\lambda_{\rm rest}$ and $f$-values for $B-X$ and $C-X$ transitions
of H$_2$* were downloaded from the MOLAT
website\footnote{http://molat.obspm.fr/} (E. Roueff, personal communication).
Table 1 lists our H$_2$* identifications for 25 absorption features
detected at the $\geq$4-$\sigma$ level in the R1200 spectrum. 

Once we were convinced of the presence of (static) H$_2$* absorption in the
data toward GRB 080607, we shifted to full dynamic modeling, with
photoexcitation, photodissociation, and photoionization of the gas
treated following DH02.  The UV and X-ray
emission from GRB 080607 was assumed to be described by
\begin{equation}
\nu L_\nu = L_0 \left(\frac{h\nu}{13.6\,{\rm eV}}\right)^{1+\beta}
                \frac{4(t/t_0)^2}{[1+(t/t_0)^2]^2}
\end{equation}
where $L_0=5\times10^{50}$ erg s$^{-1}$,
$\beta=-0.5$, and $t_0=10$ s.  With $A_V=3.2$ mag (P09), this reproduces
the observed K-band flux density at $t_{\rm obs}=300$ s.
The value of $L_0$ is 200 times higher than that adopted in DH02.
Indeed, GRB 080607 ranks among the most powerful to date, an equal
to GRB 990123 in $\gamma$ luminosity \citep{Gole08}.

Total line of sight abundances for H and H$_2$ were taken from P09,
with the H nucleon density ($n_{\rm H}$) taken to be
10$^3$ cm$^{-3}$.
The dust/gas ratio in the pre-GRB gas was assumed to be $\sim$10\%
of the local Milky Way value (P09).  For photoexcitation of the H$_2$ and
photoionization of H and H$_2$, we were primarily interested in the
dust extinction cross section per H nucleon at $\lambda < 1110$ \AA,
which we take to be $\sigma = 2\times10^{-22}$ cm$^{-2}$, so that
$A_{1000 \rm \AA} = 5.33\times10^{22}\times 2\times10^{-22}/1.086 =
9.8$ mag, or $A_{1000 \rm \AA}/A_V=3.1$.

Finally, the evolution of the irradiated dust was followed with the same
assumptions
as in DH02.  Initial grain size was $a=0.04$ $\micron$, which could decrease
by thermal sublimation if the grains became hot enough.
For calculating UV extinction we used $Q_{\rm ext}=2$,
where $Q_{\rm ext}$ is the ratio of the extinction cross section to the
geometric cross section, $\pi a^2$.

\begin{deluxetable}{cccccc}[ht]
\tablecolumns{6}
\tablewidth{0pt}
\tabletypesize{\footnotesize}
\tablecaption{H$_2$* Features Identified toward GRB 080607\label{tab:lines}}
\tablehead{ $\lambda_{\rm obs}$ & $W_{\lambda}$\tablenotemark{a} & ID in P09\tablenotemark{b} & $\lambda_{\rm rest}$ & ID in DH02\tablenotemark{c} & New ID\tablenotemark{d}\\
(\AA) &(\AA) & &(\AA) & &($v''$; $J''$) }
\startdata
5762.6  &0.8   &\nodata			&1427.1  &\nodata &(6; 0,1) \\
5783.5  &1.1   &\nodata			&1432.3  &\nodata &(6; 2) \\
5789.7  &0.6   &\nodata			&1433.8  &\nodata &(6; 4) \\
5797.4  &1.4   &\nodata			&1435.7  &P1+R1	  &(8; 0) \\
5988.6  &1.0   &\nodata			&1483.1  &\nodata &(10; 2) \\
6002.3  &2.6   &\nodata			&1486.5  &R0+R1	  &(7; 0,1) \\
6012.5  &1.9   &\nodata			&1489.0  &\nodata &(7; 1,2) \\
6115.3  &1.1   &UID			&1514.5	 &\nodata &(8; 0,1) \\
6124.8  &1.5   &UID			&1516.8	 &R1+P1	  &(7; 2) \\
6130.3  &1.3   &UID			&1518.2	 &P1	  &(11; 0,1) \\
6135.6  &0.9   &\nodata			&1519.5  &\nodata &(8; 2) \\
6156.6  &0.9   &w/ \ion{Si}{2} &1524.7  &\nodata &(7; 3) \\
6179.2  &0.6   &\nodata			&1530.3  &\nodata &(10; 1) \\
6205.8  &0.6   &\nodata			&1536.9  &\nodata &(10; 3) \\
6213.8  &0.6   &\nodata			&1538.9  &\nodata &(9; 0,1) \\
6263.7  &1.4   &w/ \ion{C}{4}	&1551.2  &\nodata &(12; 1,2) \\
6269.1  &0.6   &UID			&1552.6	 &\nodata &(13; 1) \\
6318.8  &0.7   &UID			&1564.9	 &\nodata &(13; 1) \\
6414.3  &0.9   &w/ \ion{Si}{1} &1588.6  &R0+R1	  &(10; 0,1) \\
6432.1  &0.7   &UID			&1593.0	 &\nodata &(10; 2) \\
6477.4  &1.6   &UID			&1604.2	 &\nodata &(9; 1) \\
6489.1  &1.5   &w/ \ion{Fe}{2} &1607.1  &P1+R3	  &(9; 2) \\
6525.8  &0.6   &\nodata			&1616.2  &\nodata &(12; 5) \\
6604.4  &0.7   &w/ \ion{Fe}{2} &1635.7  &\nodata &(9; 1) \\
6614.6  &1.6   &w/ \ion{Fe}{2} &1638.2  &R1+R2	  &(11; 2) \\
\enddata
\tablenotetext{a}{Detections with $W_{\lambda}$ $\geq$ 0.6 \AA\ ($\geq$4$\sigma$).}
\tablenotetext{b}{UID: Listed in P09 without ID. Otherwise, was included within a stronger atomic feature.}
\tablenotetext{c}{Strongest transitions within 1.0 \AA\ of $\lambda_{\rm rest}$ for $R$ = 350.}
\tablenotetext{d}{Strongest contributing ro-vibrational levels for $R$ = 4000.}
\end{deluxetable}

\section{Results and Discussion}

\subsection{H$_2$* Column Density}

Fig.~\ref{fig:r1200_zoom} shows a small portion of the R1200 data, superposed
by 336 modeled transitions of H$_2$*.
This region includes five absorption ``lines'' that were listed, but not
identified, in P09, four of them as a group between
$\lambda_{\rm obs}$ = 6115--6130 \AA.
However, each H$_2$* ``line'' is a blended feature that includes several strong
transitions.
It is evident that the four observed features near 6120 \AA\ are too strong
to be attributed to noise: their equivalent width ($W_{\lambda}$) values
listed in P09 range over 1.1--2.3 \AA, or 7--15 times the 1-$\sigma$
detection limit.
Since these ``lines'' cannot be identified with any other plausible atomic
or molecular species, at the redshift of either GRB 080607 or the
two intervening absorption systems, we are confident that H$_2$* provides
the only viable model for their presence in our data.
For the best match with the data, the static modeling with Ismod.f returned
log $N$(H$_2$*) = 17.7 $\pm$ 0.3, using $b$ = 2 km s$^{-1}$,
a value taken from the CO analysis of P09.
Our full dynamic modeling with the DH02 code, computing H$_2$ excitation for
10$^3$ s in the GRB time frame and using $b$ = 3 km s$^{-1}$, returned
log $N$(H$_2$*) = 17.5 $\pm$ 0.2, in good agreement with the static model.
This served to confirm that the first case of H$_2$*
excitation by GRB photon pumping has been identified.

\begin{figure}
\begin{center}
\includegraphics[width=3.5in]{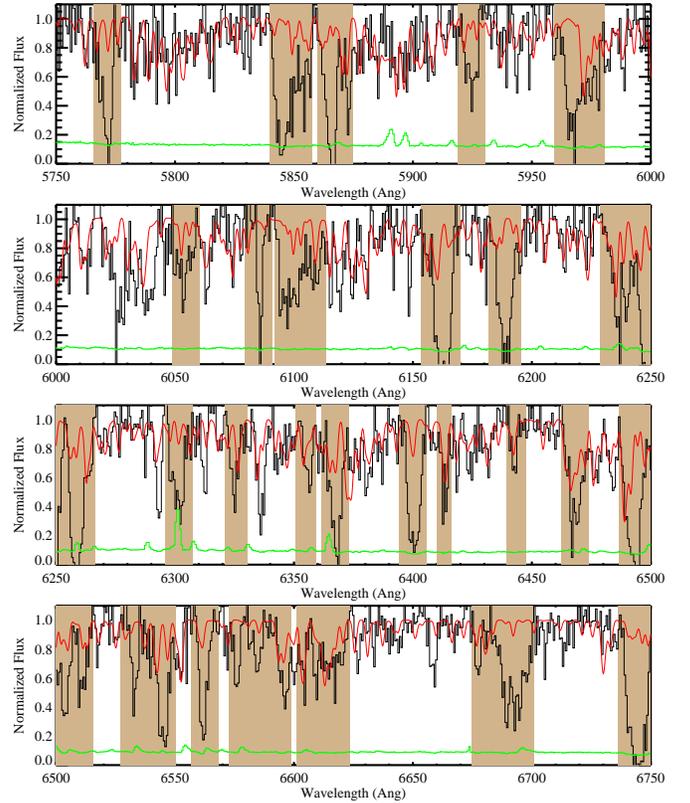}
\end{center}
\caption{Same as Fig.~\ref{fig:r1200_zoom}, but showing 1000 \AA\ of the
R1200 data and the model of H$_2$*.
Dozens of narrow H$_2$* and observed features appear to agree with each other.
The global pattern of variable $v'-v''$ band strength is also matched.}
\label{fig:r1200}
\end{figure}

The entire range of the R1200 spectrum ($\lambda_{\rm rest}$ = 1400--1720 \AA)
covers the red portion of H$_2$* absorption, and includes 3372 transitions with
$J''$ = 0--29, $v''$ = 1--14, and $v'$ = 0--16.
Figure~\ref{fig:r1200} presents a global view of the H$_2$* model 
spectrum superposed over 1000 \AA\ of the R1200 data.
The overall agreement between data and model is acceptable, given that
the S/N $\approx$ 10 and the presence of numerous other absorption
features from, e.g., host-galaxy CO bandheads and \ion{Fe}{2} from
two foreground absorbers (P09).
The few instances where the modeled absorption is significantly deeper
than the data involve levels with
$v''$ $\geq$ 9, presumably related to our incomplete description
(underestimation) of UV photoionization rates out of high-$v''$ levels.

We identified 25 narrow H$_2$* features that are the least blended with other
species and list these in Table~\ref{tab:lines}.
Previously, P09 estimated atomic abundances in the weak limit from
measured $W_{\lambda}$ of absorption features.
Since some transitions listed in P09 are seen here (Fig.~\ref{fig:r1200})
to be blended with some of the stronger H$_2$* features,
certain $W_{\lambda}$ values should be revised downward.
For example, the H$_2$* feature at 6466 \AA\ falls into a blend with the
\ion{Ge}{2} 1602 line, contributing 2.0 \AA\ to the total $W_{\lambda}$ of
5.3 \AA\ and reducing its weak-limit abundance (estimated by P09 to be
super-solar).
However, only a future detailed study of sight-line saturation will be able
to provide robust abundance determinations based on H$_2$*-corrected data.

The presence of H$_2$* in our R1200 spectrum is confirmed
by the lower-S/N exposures from the B600 grating ($R$ = 2000).
In all, the blue and red gratings provide $\sim$500 \AA\ of rest-frame
coverage of numerous absorption
features belonging to (or blended with) H$_2$* toward GRB 080607. 

\begin{figure}
\begin{center}
\includegraphics[width=3.5in]{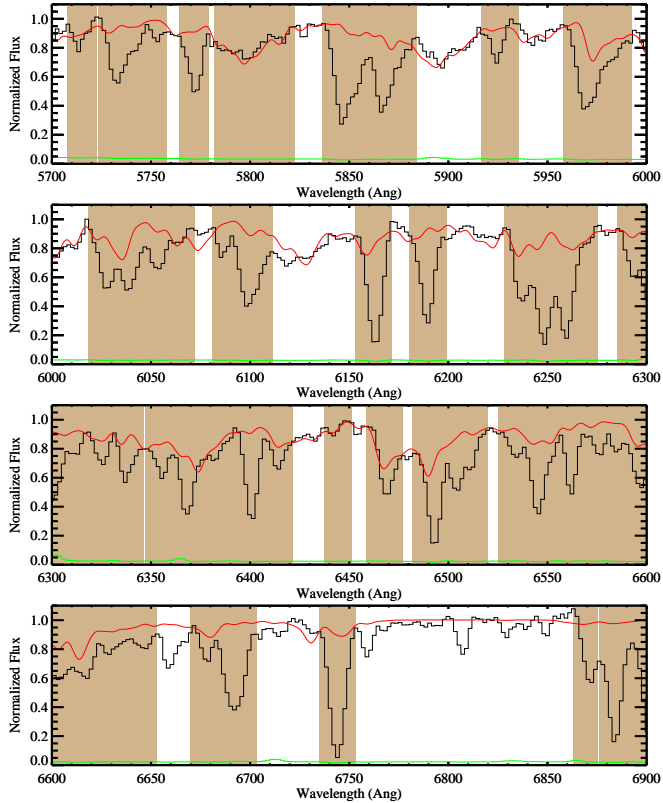}
\end{center}
\caption{Same as Fig.~\ref{fig:r1200}, but showing the lower-resolution
R400 data and H$_2$* model. 
Absorption from wide features made up of numerous H$_2$* transitions is
determining the shape of the continuum of the R400 spectrum.
Some absorption features previously assigned to atomic carriers are clearly
significantly blended with contribution from H$_2$*.}
\label{fig:r400}
\end{figure}

\subsection{H$_2$* Variability vs. Non-Variability}

The theoretical work of DH02 followed the temporal behavior of H$_2$*
abundance in response to the initial UV flash from a GRB afterglow.
Upon photoexcitation into the electronic $B$ and $C$ states, only 
$\sim$15\% of H$_2$ is photodissociated, 
with the majority, therefore, decaying immediately
into $v > 0$ levels of the $X$ ground state.
A very rapid rise in $N$(H$_2$*) occurs over the time scale of the burst
($\approx$10 s) as the UV photons get absorbed by the cold H$_2$ gas.
The 10$^6$ s lifetime of H$_2$* \citep{Drai00}
means that it remains around for almost two months in the observer frame.

As described in \S 2, our two groups of red exposures were centered 
on $t$ $\sim$ 500 and 1400 s in the GRB rest frame.
As shown in Fig.~\ref{fig:r400}, a degraded $R$ = 1000 version
of the H$_2$* model that matches the R1200 spectrum shows remarkable
agreement with the upper envelope of the R400 spectrum.
Thus there is no indication of variation in $N$(H$_2$*) over the
$\Delta$$t$ $\sim$ 900 s that elapsed between the clusters of R400 and
R1200 exposures.
This observational result confirms the DH02 theoretical predictions that
$N$(H$_2$*) varies over time scales that are either significantly shorter
or significantly longer than those covered by our observations.

\subsection{The Distance of H$_2$* from the GRB}

The value of $N$(H$_2$*) detected here is only $\sim$3\% of that
produced by the models in DH02.
According to DH02, the amount of H$_2$* produced depends
on the fluence, $F$, of the source, and hence on the ratio $L_0/D^2$,
where $D$ is the distance from the GRB to the center of the initial H$_2$
integration zone.
Owing to the much higher value of $L_0$ used in our modeling of GRB 080607
($\S2$), the production of an appreciably
{\it smaller} amount of H$_2$* relative to DH02 would require a very large
increase in $D$.

We employ the detailed modeling of DH02 in order to explore the dependence
of $N$(H$_2$*) on $D$.
The effect of the GRB on the H$_2$ will depend on what fraction of the
\ion{H}{1}
is located between the GRB and the H$_2$, because the dust in the \ion{H}{1}
gas can help shield the H$_2$ from photons with $\lambda <1110$ \AA,
and the \ion{H}{1} itself will help protect the H$_2$ from photoionization.

We consider two cases.  Model A has $N({\rm H^0})=2\times10^{21}$
cm$^{-2}$ located between the GRB and the H$_2$, with the
remainder of the \ion{H}{1} located beyond the H$_2$.  Model B has $N({\rm
  H}^0)=5\times10^{22}$ cm$^{-2}$ (i.e., {\it all} of the \ion{H}{1} from P09)
located between the GRB and the H$_2$.  For a given large value of
$D$, model B will have less UV-pumped H$_2$ than model A, because of
the additional dust shielding in model B.  However, for smaller values
of $D$, reduced shielding by the destruction of dust and \ion{H}{1}
in model B may result in
complete destruction of the H$_2$ by photodissociation or
photoionization.

\begin{figure}
\begin{center}
\includegraphics[width=3.5in]{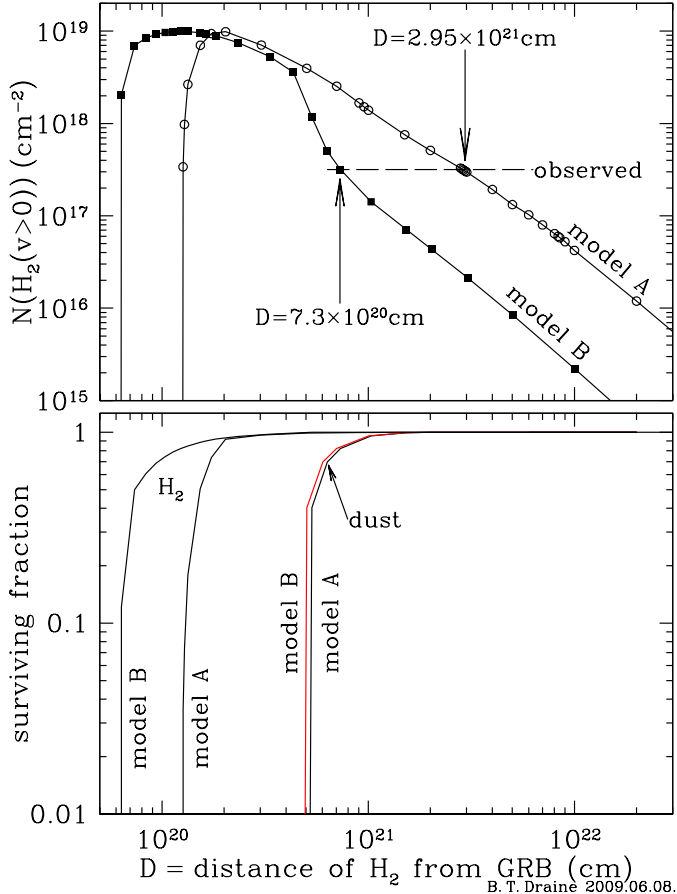}
\end{center}
\caption{\label{fig:models}Column density of vibrationally-excited H$_2$
as a function of distance from GRB 080607, for two models differing in
the location of the \ion{H}{1} relative to the H$_2$ (see text).}
\end{figure}

The upper panel of Fig. \ref{fig:models} shows
$N$(H$_2$*) versus $D$ for the two models.
The observed value of $\approx 10^{17.5}$ cm$^{-2}$ ($\S$3.1)
is generated for $D=2.9\times10^{21}$ or $7.3\times10^{20}$ cm
for Model A or B, respectively.
Thus we conclude that the H$_2$ in GRB 080607 is located between
230--940 pc from GRB 080607, a value much larger than that used
in DH02 ($D$ $\ll$ 1 pc).

The lower panel of Fig. \ref{fig:models} shows the fraction of the
initial dust and H$_2$, that survives after the GRB flash.
For more than
50\% of the initial grain mass to be sublimed, the H$_2$ gas would have to be
within $\sim$190~pc, slightly closer than the location of the H$_2$* in
model B, and substantially closer than in model A.  We conclude that for
these two models, the bulk of the dust survives the GRB flash, consistent
with $A_V\approx3.2$ mag observed during the afterglow phase.
Substantial destruction of the H$_2$ by a combination of
photodissociation and photoionization occurs only for $D <
2\times10^{20}$ cm $=65$ pc. Given the observed
$N$(H$_2$*)/$N({\rm H}_2)\approx10^{-3.7}$, we
conclude that there has been minimal destruction of H$_2$ in the cloud
where the H$_2$* resides.

More complex models are of course possible: if more than one H$_2$
cloud is present, there could have near-complete destruction of the
H$_2$ in the inner cloud, leaving only a small amount of H$_2$ with
high levels of vibrational excitation, with the bulk of the H$_2$ in a
more distant cloud with little vibrational excitation.

\subsection{What about CO*?} 

The presence of H$_2$* prompted us to search for vibrationally-excited CO
(CO*) in our data.
P09 reported the detection of log $N$ = 16.5 $\pm$ 0.3 cm$^{-2}$ of
cold ($v''$ = 0) CO, all of which was rotationally-excited (hereafter
r-CO*) at $T_{\rm ex}$ $\sim$ 300 K up to $J''$ = 25.
If the vibrational excitation of CO owing to UV pumping scales with that
of H$_2$, then we should expect
log $N$(CO*) = log $N$(CO) $-$ 3.7 $\approx$ 12.8 cm$^{-2}$.
CO* transitions for $v''>0$ bands were taken from Kurucz line
lists\footnote{http://kurucz.harvard.edu/LINELISTS/LINESMOL/}
and modeled with Ismod.f, but no match with the R1200 data was found.
Based on the strongest $v''=1$ band at 6400 \AA, a feature with
$N$(CO*) = 1 $\times$ 10$^{13}$ cm$^{-2}$ would have a depth of 10\%,
comparable to the 1-$\sigma$ noise level in the data.

We note that the photophysical behavior of CO differs from that of H$_2$
in two important respects.
First, owing to the existence of strongly predissociating Rydberg states
in CO below $\lambda_{\rm rest}$ = 1100 \AA\ \citep{Letz87,vanD88,Shef03},
CO*/CO should be much lower than H$_2$*/H$_2$.
Second, CO* spontaneously decays via dipole transitions with a lifetime of
ca. 10$^{-2}$ s \citep{Okad02}, roughly 10$^8$ times faster than the
quadrupole transitions of H$_2$*.
Any small amount of surviving CO* would be rapidly converted into r-CO*.
However, while CO photophysics may explain the absence of CO* toward GRB 080607,
the challenge of accounting for the conversion of the bulk of cold CO
into r-CO* remains.

Both dust extinction and cold H$_2$ provide very effective
UV shielding of CO below $\lambda_{\rm rest}$ = 1100 \AA,
allowing it to survive in deeper layers of a cloud.
The non-dissociating $A-X$ bands (with $v'$ $\leq$ 8) are
significantly less shielded:
the data toward GRB 080706 (Fig. 1 of P09) show that the UV flux level
at $\lambda_{\rm rest}$ $>$ 1322 \AA\ is $\sim$5 times higher than
at $\lambda_{\rm rest}$ $<$ 1150 \AA.
UV pumping into the CO $A$ state could result in
its complete conversion into CO* and then into r-CO*.
However, assuming that initially the CO is cold
($T_{\rm ex} \sim$ 11 K, $J''$ $\leq$ 5), the production of r-CO* that
happens to maintain a quasi-thermal distribution at $\sim$300 K up to
$J''$ = 25 must be considered a coincidence of indirect excitation.

Radio observations show that collisionally-excited r-CO* can be found in
Galactic PDRs \citep{Tiel85}, with $T_{\rm ex}$ $\sim$ 100--1000 K
\citep{vanD88,Drai96,Holl97}.
This range agrees with $T_{\rm ex}$ $\sim$ 300 K for r-CO* toward GRB 080607,
subject to the caveat that appreciably higher values of $N$(CO) and $A_V$
are found in Galactic PDRs.
We surmise that the r-CO* toward GRB 080607 could be a pre-burst observational
signature of the PDR produced by the progenitor of the GRB, as it was
forming a giant \ion{H}{2} region around itself \citep{Whal08}.
A detailed calculation of this process is clearly warranted.

\section{Concluding Remarks}

In addition to providing the first positive detection of H$_2$ bands
and the first observation of CO bands in a GRB-host galaxy (P09), we have
shown that GRB 080607 also provides the first evidence for H$_2$*
in a GRB host galaxy and marks the highest-redshifted H$_2$* detected to date.

This discovery of UV-pumped H$_2$* toward GRB 080607 serves to confirm
the predictions for its production under precisely such circumstances
\citep[][DH02]{Drai00}.
Our initial static modeling with Ismod.f indicated log $N$(H$_2$*) $<$ 18.0
toward GRB 080607, significantly lower than the original predictions of DH02.
We then showed that the dynamic models of DH02 can successfully reproduce
observed H$_2$* absorption once modifications involving individual
GRB luminosities and adjustable distance to the molecular cloud are
incorporated.
Thus the DH02 photoexcitation code shows that the bulk of the molecular
cloud harboring H$_2$* is located at a model distance of 230--940 pc
along the line of sight from the UV afterglow.
This is much farther than the original arrangement in DH02, but still in
the local galactic neighborhood of GRB 080607.
In the scheme of \citet{Whal08}, a model distance of $>$230 pc for the molecular
cloud means that it is located outside the $\approx$100 pc radius of
\ion{H}{2} region carved by the progenitor of GRB 080607.

One interesting result of the increased distance from the GRB to the gas
is the inability of the radiation beam to destroy dust embedded in the cloud.
Whereas in DH02 dust was easily destroyed at close quarters
to the GRB, following its heating to 2000--3000 K, this process cannot operate
at larger distances, leading to survival of the dust in the cloud. 
Such dust remains on the line of sight between the GRB and the H$_2$
destruction front, diminishing the effects of such destruction.

\acknowledgements
Y.S. is partially supported by S. Federman, who suggested the
production of rotationally-excited CO in a PDR.
J.X.P. is partially supported by NASA/Swift grants
NNG06GJ07G and NNX07AE94G and an NSF CAREER grant (AST-0548180).
B.T.D. is partially supported by NSF grant AST-0406883.
D.A.P. is partially supported by HST grant HST-GO-11551.01-A.
J.S.B. is partially supported by an Alfred P. Sloan Foundation fellowship.

\end{document}